\documentclass[
  a4paper, 
  12pt, onecolumn,
]{article}

\title{\bf   Derivation  of a \\ Matrix Product Representation
                for the Asymmetric Exclusion Process from
                       Algebraic Bethe Ansatz
}
\author{        O. Golinelli, K. Mallick
\\ \normalsize\em          
Service de Physique Th\'eorique, Cea Saclay, 91191 Gif, France
}
\date{\normalsize  April 13, 2006 \\ 
arXiv:cond-mat/0604338
}

\begin{document}
\maketitle

\begin{abstract}
 
   We derive,  using the  algebraic Bethe Ansatz, a 
 generalized  Matrix Product  Ansatz  for 
 the  asymmetric exclusion process (ASEP) on a
  one-dimensional  periodic lattice.
  In this Matrix Product  Ansatz, 
   the components of  the  eigenvectors  of the ASEP Markov 
 matrix can be expressed  as traces of  products 
 of non-commuting operators. 
  We derive  the relations between the operators 
  involved  and show that they generate a quadratic  algebra.
  Our construction provides explicit
  finite dimensional  representations
  for the  generators of this algebra.  
  \normalsize

  \medskip \noindent Keywords: ASEP, Matrix Product Ansatz,
  Algebraic  Bethe Ansatz.
  
  \medskip \noindent Pacs number: 05.40.-a, 05.60.-k

\end{abstract}

\section{Introduction}

The asymmetric simple exclusion process (ASEP) 
 that   plays a fundamental role in the theoretical studies
 of  non-equilibrium statistical mechanics, 
 is a driven lattice gas model in which particles
  interact by hard core exclusion.  The ASEP was originally 
   introduced as a building block for models
  of one dimensional transport where geometric constraints
  play an important  role  (e.g., hopping
  conductivity, motion of RNA templates and  traffic flow).
   
 The  exclusion process is a stochastic  Markovian  model 
 whose dynamical  rules are  encoded in an 
 evolution (Markov) matrix. 
 Exact results for the ASEP in one dimension   have been derived using two
  complementary approaches, the Matrix Product Ansatz and the Bethe Ansatz 
  (for a review see Derrida 1998, Sch\"utz 2001). 
  The Matrix Product Ansatz (MPA)
  (Derrida {\it et al.} 1993)   is based on 
  a representation  of the components of the steady  state wave function
 of the Markov operator  in terms of a product of matrices. 
  This method  has been used   to calculate steady state
 properties  of the ASEP such as the invariant measure (Speer 1993), 
 current fluctuations in
 the stationary state and large 
 deviation functionals (Derrida {\it et al.} 2003). 
 
 The ASEP is equivalent to a non-hermitian Heisenberg 
 spin chain of the XXZ type
 and  can  be mapped into a two-dimensional six-vertex model 
 at equilibrium~:  these mappings permit the  use of  integrable
 systems techniques, such as the  Bethe Ansatz.
 The Bethe Ansatz provides spectral information about the
  evolution operator (Dhar 1987, Gwa and Spohn 1992; 
 Kim 1995,  Golinelli and Mallick 2004)
  which can  then be used to derive large deviation
  functions (Derrida and Lebowitz  1998).

 The exact relation between these two techniques has been 
  a matter of investigation  for a long time (Alcaraz
{\it et al.} 1994; Stinchcombe and Sch\"utz 1995).
 In a recent work,  Alcaraz and Lazo (2004)  have expressed 
 the  eigenvectors of   integrable 
  quantum chains (such as the anisotropic Heisenberg chain)
  as traces of products of generators of   a quadratic algebra.  This 
  Matrix Product Ansatz leads to  the 
   Bethe Ansatz equations of the system. 

  In this work, we  solve the inverse problem:  we prove   that a  
  Matrix Product representation  involving 
   quadratic algebraic relations  between  operators 
  can be {\it deduced and   constructed  explicitly} 
  by    applying    the Algebraic  Bethe Ansatz  to the ASEP.
  The  quadratic  algebra  we obtain  is similar to the one 
  studied by  Alcaraz and Lazo. However, our  algebra 
  satisfies different boundary conditions that   modify  
   drastically the properties of its representations and ensure the  
   existence of   finite dimensional representations.

 The outline of this work is as follows. In section \ref{sec:ABA}, 
 we apply  the Algebraic Bethe Ansatz  to  the 
 totally asymmetric exclusion process.
 In section \ref{sec:MPA}, we derive the Matrix Product Ansatz from the
 Algebraic Bethe Ansatz. In section \ref{sec:quad}, we establish 
 the quadratic algebraic relations  satisfied by the operators of  the MPA. 
 Our results are generalized to the 
 partially asymmetric  exclusion process  in  Section \ref{sec:PASEP}.
 Concluding remarks
 are presented in the last section. In the appendix, we
 derive the  Bethe Ansatz equations from  the quadratic algebra
 of section  \ref{sec:quad}.

\section{Algebraic Bethe Ansatz for ASEP}
\label{sec:ABA}

We consider the exclusion process   on a periodic
one dimensional lattice with $L$ sites (sites $i$ and $L + i$ are
identical). 
 A  lattice site cannot be  occupied by more than  one particle.
 To represent the state of a site, $i$ ($1 \le i \le L$)  we
 use  the spin-1/2 language~:  a site $i$
   can be in two states that we  label
 as  $ |\uparrow\rangle$ ($i$ is occupied) and 
 $ |\downarrow\rangle$ ($i$ is empty). 
 A configuration ${C}$ is  represented  either  by 
 a vector of the type  $| \downarrow \uparrow \ldots \uparrow \rangle$
 or by a binary  vector
\begin{equation}
  |{C} \rangle =  |\tau_1,\ldots,\tau_L \rangle \, .
  \label{eq:Ctau}
\end{equation}
 where  $\tau_i  = 1$ if   
the   site $i$ is occupied  and    $\tau_i  = 0$  otherwise.
  The space of all  possible configurations
  is a     $2^L$ dimensional vector space
 that we shall denote by   $\mathcal{S}$.

 The system evolves   with  time according to 
the following stochastic rule: a particle on a site $i$ at time $t$ jumps, in
the interval between  $t$ and $t+dt$, with probability $p\ dt$ to the
neighbouring site $i+1$ if this site is empty ({\em exclusion rule})
 and with  probability $q\ dt$ to the
 site $i-1$ if this site is empty. The jump  rates $p$ and $q$
 are normalized such that  $ p + q =1$. In 
  the totally asymmetric exclusion process (TASEP),  
 the jumps are totally biased in one direction ($p =1$ and $q =0$).
 For  sake of simplicity, we shall discuss the TASEP case in full detail.
 The general case will be considered briefly  in Section \ref{sec:PASEP}.

 We call $\psi_t({C})$
 the probability of a  configuration ${C}$ at time $t$.
  As the exclusion process is a continuous-time Markov process, the time
 evolution of $\psi_t({C})$ is determined by the master equation 
\begin{equation}
    \frac{d}{dt} \psi_t({C})  = \sum_{{C}'}
      M({C},{C}') \psi_t({C}')  \, , 
\end{equation}
  where  the element  $ M({C},{C}')$
 is the  transition rate  from configuration ${C}'$ to ${C}$
 and  the diagonal term $M({C},{C}) =
  -  \sum_{{C}'}  M({C}',{C})$ 
represents  the exit rate  from configuration ${C}$.
 The Markov  matrix  $M$, that  encodes  the dynamics
  of the exclusion process,
 is   a  square matrix of size  $2^L$  acting  on the configuration
 space  $\mathcal{S}$ of the TASEP.
 This Markov matrix can be expressed
 as a sum of local  operators that  update  the
 bond $(i, i+1)$~:
\begin{equation}
             M =  \sum_{i=1}^{L} M_{i, i+1}  \, ,  
\label{eq:sumlocalM}
\end{equation}
 where  the   TASEP local  update operator  $ M_{i, i+1}$ is given by
\begin{equation}
         M_{i, i+1}  = 
 {\bf 1}_1 \otimes {\bf 1}_2 \ldots {\bf 1}_{i-1}  \otimes
 \left( \begin{array}{cccc}
                         0 & \,\,\,\,  0&  0 &  0  \\
                         0 & - 1&  0  &  0  \\
                         0 & \,\,\,\, 1&  0  &  0  \\
                         0 & \,\,\,\,  0&  0 &  0  
                   \end{array}
                    \right)  
         \otimes {\bf 1}_{i+2} \ldots {\bf 1}_L \, .
\label{eq:localM}
\end{equation}
 The matrix   ${\bf 1}_{j}$ is the $2 \times 2$ identity matrix
 acting on the site  $j$ and    the $4  \times 4$ matrix 
 appearing in this equation is written in the
  local  basis  ($|\uparrow\uparrow~\rangle,$ 
 $|\uparrow\downarrow~\rangle,$  $|\downarrow\uparrow~\rangle,$ 
 $|\downarrow\downarrow~\rangle$ ),
 that represents the four possible states of the 
   bond $(i, i+1)$. Thus,   $ M_{i, i+1}$ is   a  $2^L$ matrix
 that acts trivially on all sites other than  $i$ and $i+1$
 and makes a particle jump from site $i$ to the site $i+1$.

 It is also useful to define the permutation operator
 $P_{i, i+1}$  that exchanges the states of sites  $i$ and   $i+1$:
\begin{equation}
         P_{i, i+1}  = 
 {\bf 1}_1 \otimes {\bf 1}_2 \ldots {\bf 1}_{i-1}  \otimes
 \left( \begin{array}{cccc}
                         1 &    0&  0  &  0  \\
                         0 &    0&  1  &  0  \\
                         0 &    1&  0  &  0  \\
                         0 &    0&  0  &  1  
                   \end{array}
                   \right)  
         \otimes {\bf 1}_{i+2} \ldots {\bf 1}_L \, .
\label{eq:localP}
\end{equation}

  More generally, we can define  a  jump  operator $M_{i, j}$ 
  and a  permutation  operator  $P_{i, j}$, 
  between sites $i$ and $j$ that act 
trivially on all sites other than  $i$ and $j$;   $ M_{i, j}$ 
  makes a particle jump from site $i$ to  site $j$ and
$P_{i, j}$  exchanges the states of sites  $i$ and  $j$. 
 
 In the Algebraic Bethe Ansatz method 
  (see e.g.,  Nepomechie, 1999  for an  introduction to this subject), 
          an auxiliary site  $a$  is introduced, 
  which can be in two states  labelled 
 as  $ |1\rangle$  (site $a$ is occupied)  and  $ |2\rangle$
   (site $a$ is empty).   These
 two states  span  a two dimensional vector space, ${\mathcal A}$,
 the auxiliary space.  We now define an  operator 
  $ {\mathcal L}_{i}(\lambda)$   that acts   on the tensor space 
 ${\mathcal A} \otimes {\mathcal S}$;  this operator  
  acts trivially on all sites other than  $a$ and $i$, and  is a function of 
 a  spectral parameter $\lambda$ 
\begin{equation}
  {\mathcal L}_{i}(\lambda)  = 
   P_{i, a} \left( 1  + \lambda M_{i, a} \right)  \, .
 \label{eq:def1Li}
\end{equation}
This operator can  also be  represented   as  a  $2 \times 2$
  operator  on   the vector space ${\mathcal A}$, 
\begin{equation} 
     {\mathcal L}_{i}(\lambda)  =   \left( \begin{array}{cc}
                   a(\lambda) & b(\lambda)  \\
                   c(\lambda) & d(\lambda)  
                         \end{array}
                    \right)     \, , 
\label{eq:defLi}
 \end{equation}
   where  the  matrix elements
  $a(\lambda),  b(\lambda),  c(\lambda)$  and $d(\lambda)$ 
  are themselves  $2^L \times 2^L$  operators
 that act    on  the configuration space  ${\mathcal S}$.
  These operators    act trivially on all sites
 different from $i$ and  are given by 
\begin{eqnarray} 
 a(\lambda) &=& {\bf 1}_1 \otimes  \ldots {\bf 1}_{i-1}\otimes
             \left( \begin{array}{cc}  1 & 0 \\
                                       0 & 0   \end{array} \right) 
      \otimes {\bf 1}_{i+1} \ldots {\bf 1}_{L} \, ,\label{eq:defaL} \\ 
b(\lambda)  &=& {\bf 1}_1 \otimes \ldots {\bf 1}_{i-1}\otimes
              \left( \begin{array}{cc}  0 & 0  \\
                                      1 - \lambda & 0   \end{array} \right)
 \otimes {\bf 1}_{i+1} \ldots {\bf 1}_{L}  \, ,\label{eq:defbL} \\
 c(\lambda)  &=& {\bf 1}_1 \otimes \ldots {\bf 1}_{i-1}\otimes
                       \left( \begin{array}{cc}  0 & 1 \\
                                       0 & 0   \end{array} \right) 
   \otimes {\bf 1}_{i+1} \ldots {\bf 1}_{L}  \, ,\label{eq:defcL} \\
d(\lambda)  &=& {\bf 1}_1 \otimes \ldots {\bf 1}_{i-1}\otimes
           \left( \begin{array}{cc}  \lambda   & 0 \\
                                   0 &       1  \end{array} \right) 
      \otimes {\bf 1}_{i+1} \ldots {\bf 1}_{L} \label{eq:defdL}  \, . 
\end{eqnarray}
  The tensor products  in these  equations   represent
  products over  the two-dimensional local  configuration space 
  of a site. 

   The operators  $ {\mathcal L}_{i}(\lambda)$  satisfy a  
   Yang-Baxter type relation.  We   consider the operator
\begin{equation}
  {\mathcal R}(\nu) = 1 + \nu M_{a', a}  \, ,
\end{equation}  
  that acts   on  ${\mathcal A} \otimes {\mathcal A}'$
  where  the  auxiliary
 spaces ${\mathcal A}$ and ${\mathcal A}'$  correspond
  to the  auxiliary sites $a$ and $a'$.   
 In   the basis   $ \left( |1_a, 1_{a'}\rangle,|1_a, 2_{a'}\rangle,
   |2_a, 1_{a'}\rangle, |2_a, 2_{a'}\rangle \right)$ of 
  ${\mathcal A} \otimes {\mathcal A}'$, the 
  operator ${\mathcal R}(\nu)$   is  
 represented by a $4 \times 4$ scalar  matrix~: 
\begin{equation}
 {\mathcal R}(\nu)  =    \left( \begin{array}{cccc}
  1 & 0 & 0 & 0 \\
  0 & 1 & \nu & 0 \\
  0 & 0 & 1 - \nu & 0 \\
  0 & 0 & 0 & 1
        \end{array} \right)     \, .     
\label{eq:MatR}
\end{equation}
  The following identity is then satisfied 
\begin{equation}
 {\mathcal R}(\nu)
   \big[ {\mathcal L}_{i}(\lambda) \otimes {\mathcal L'}_{i}(\mu)
    \big]
=   \big[  {\mathcal L}_{i}(\mu) \otimes {\mathcal L'}_{i}(\lambda)  \big]
    {\mathcal R}(\nu) \,\,\, {\rm with }  \,\,\,
  \nu =  \frac{\lambda - \mu}{ 1 - \mu} \, , 
\label{eq:YBE}
\end{equation}
 where  ${\mathcal L}_{i}$ and ${\mathcal L'}_{i}$  
 are interpreted as  $2 \times 2$  matrices
  acting  on  ${\mathcal A} $  and ${\mathcal A'}$, respectively, 
  with matrix elements that are themselves operators on ${\mathcal S}$.
 Their tensor product is thus a $4 \times 4$  matrix,
 acting  on  ${\mathcal A}\otimes {\mathcal A'}$ 
 with  matrix elements that are operators on ${\mathcal S}$.

 The  {\it monodromy}  matrix is defined as
\begin{equation}
  T(\lambda) =  {\mathcal L}_{1}(\lambda) {\mathcal L}_{2}(\lambda)
 \ldots  {\mathcal L}_{L}(\lambda) 
 \, ,
\label{eq:defTlambda}
\end{equation}
 where the product of the  ${\mathcal L}_{i}$'s has to be understood as
  a product of  $2 \times 2$ matrices acting on ${\mathcal A}$
 with non-commutative elements. 
 The  monodromy  matrix  $T(\lambda)$
  can thus  be written as
\begin{equation}
  T(\lambda) =  \left( \begin{array}{cc}
                      A(\lambda)  &  B (\lambda)  \\
                      C(\lambda)  &  D (\lambda) \end{array}
                    \right)     \, ,
\label{eq:def2T}
\end{equation}
where  $A, B, C$ and $D$ are operators on the configuration 
  space  ${\mathcal S}$. 
 Taking the trace of the   monodromy  matrix
 over the auxiliary space  $\mathcal{A}$, we obtain a one-parameter
 family of {\it  transfer}  matrices   acting on  ${\mathcal S}$
\begin{equation}
  t(\lambda) = 
{\rm Tr}_\mathcal{A}( T(\lambda)) =   A(\lambda) +  D (\lambda) \, .
  \label{eq:deftlam}
\end{equation}
  Equation~(\ref{eq:YBE})  implies  that the operators
  $t(\lambda)$   form a family of commuting operators
 (see e.g.,  Nepomechie, 1999). 
  In particular,  this family contains the  translation  operator
  ${\mathcal T} =  t(0)$ 
 (that shifts  all the particles   simultaneously 
  one site forward) and 
 the Markov matrix   $M = t'(0)/t(0)$. Using  Algebraic Bethe Ansatz, 
 the  common eigenvectors of this   family  are   explicitly  constructed 
  by actions  of  the  $B$ operators on
  the  reference   state $\Omega$, defined as
 \begin{equation}
       \Omega =  | \uparrow\uparrow \ldots \uparrow \rangle  \, . 
   \label{eq:defOmega}
\end{equation}
 The state   $\Omega$  corresponds to a  configuration where all the 
sites are occupied.

 More precisely,  for any  $n \le L$,   we define the vector
\begin{equation}
    | z_1, z_2 \ldots  z_n    \rangle  
 =     B(z_n) \ldots B (z_2) B (z_1)\, \Omega \,,  
\label{eq:ABA}
\end{equation}
 where $ z_1, z_2 \ldots  z_n$ are complex numbers. 
  Because  each operator $B$ creates a hole in the system, 
 the   state $| z_1, z_2 \ldots  z_n    \rangle$
  is a linear combination  of
 configurations  with exactly  $n$ holes. This vector   
 is an eigenvector of the  operator  $t(\lambda)$ 
 (for all values of $\lambda$) 
   and in particular
 of the Markov matrix M,  provided  the pseudo-moments  
$z_1$, $z_2$, $\ldots$,
   $z_n$ satisfy the Bethe equations~:
\begin{equation}
   z_l^L  = (-1)^{n-1} \prod_{i =1}^n \frac{1 -z_l}{1 -z_i} \,\,\,
  {\rm  for }\,\,\,\,  l =1 \ldots n\, .
\label{eq:Bethe}
\end{equation} 
 The corresponding eigenvalue of  $t(\lambda)$ is given by 
 \begin{equation}
     E(\lambda) = 
       \frac{ (1 -\lambda)^n  +  \lambda^L   \prod_{i =1}^n (z_i -1)}
      { \prod_{i =1}^n  (z_i -\lambda)}     \, .
 \label{eq:eigenval}
 \end{equation}
 Using the Bethe equations~(\ref{eq:Bethe}),  we find 
 that  $E(\lambda)$ is a polynomial in  $\lambda$
 of degree $L-n$.
 
\section{Derivation  of the  Matrix Product Representation from Bethe Ansatz}
\label{sec:MPA}

  In the previous section, we have constructed 
   the eigenvectors of the  Markov matrix by using the
  the Algebraic Bethe Ansatz, see equation~(\ref{eq:ABA}).
  In this section we show that
  the Algebraic Bethe Ansatz  also permits us to express   the components
  of an eigenvector as a Matrix Product (Mallick  1996).
   From  equation~(\ref{eq:def2T}), we remark  that
\begin{equation}
        B (\lambda)  =   \langle 1  |  T(\lambda) | 2  \rangle \, .
\label{eq:BfromT}
\end{equation}
 Using this relation,  the eigenvector given in equation~(\ref{eq:ABA})
 can be written as 
\begin{eqnarray}
    | z_1, z_2 \ldots  z_n    \rangle  
  &=& \langle 1  |  T(z_n) | 2  \rangle \, 
  \ldots \langle 1  |  T(z_2) | 2  \rangle 
      \langle 1  |  T(z_1) | 2  \rangle  \, \Omega   \nonumber  \\
 &=&    \langle 1, 1, \ldots, 1  |  T(z_n) \ldots  \otimes  T(z_2) \otimes
    T(z_1)  | 2, 2, \ldots, 2  \rangle  \, \Omega \nonumber  \\
   &=& {\rm Tr} \Big( { Q}_n   T(z_n) \otimes \ldots
      \otimes   T(z_1)  \, \Omega   \Big) \, , 
\label{eq:ABA2}
\end{eqnarray}
 where the tensor-products act  on the  space ${\mathcal A}^{\otimes n}$
 with $n$ auxiliary sites. 
 The boundary operator ${Q}_n$ is given by
  \begin{equation}
       { Q}_n  =   | 2, 2, \ldots, 2  \rangle 
  \langle 1, 1, \ldots, 1  |  \, .
\label{eq:defQn}
\end{equation}
 Using the definition~(\ref{eq:def2T}) of the $T$ matrix, we rewrite
 equation~(\ref{eq:ABA2}) as follows
 \begin{eqnarray}
   | z_1, z_2 \ldots ,  z_n    \rangle   &=&  
 {\rm Tr} \Big( {Q}_n  
  \prod_{i=1}^L  {\mathcal L}_{i}(z_1,\ldots,z_n) \, \Omega   \Big) \, , 
\label{eq:ABA3}    \\
\hbox{ with } \,\,\,\, 
  {\mathcal L}_{i}(z_1,\ldots,z_n)  &=&    {\mathcal L}_{i}(z_n) 
  \otimes \ldots  \otimes   {\mathcal L}_{i}(z_1) \, . 
 \label{eq:ABA4} 
\end{eqnarray}
 The operator  ${\mathcal L}_{i}(z_1,\ldots,z_n)$ is a  
  $2^n \times 2^n$ matrix  acting on ${\mathcal A}^{\otimes n}$.
  The  matrix elements of  ${\mathcal L}_{i}(z_1,\ldots,z_n)$
 are  operators on the 
 configuration space ${\mathcal S}$ that  act trivially
 on all sites except  the site $i$. 
 We now define the two  operators $D_n$ and $E_n$ by the following
 relation 
  \begin{equation}
 {\mathcal L}_{i}(z_1,\ldots,z_n) \,  |\uparrow \rangle
 =  D_n (z_1,\ldots,z_n) \, |\uparrow \rangle + 
    E_n (z_1,\ldots,z_n)  \, |\downarrow \rangle \, , 
\label{eq:defDE}
\end{equation}
 where $ |\uparrow \rangle$ and  $|\downarrow \rangle$ represent
 the states  of the site $i$.  (For sake of simplicity, we are writing
   $|\uparrow \rangle$ and  $|\downarrow  \rangle$  instead of 
  $|\uparrow_i \rangle$ and  $|\downarrow_i \rangle).$
 
 Equivalently, these two operators are given by 
\begin{eqnarray}
         D_{n}(z_1,\ldots,z_{n})  &=&
 \langle \uparrow | 
  {\mathcal L}_{i}(z_1,\ldots, z_{n} ) \,  |\uparrow \rangle
\label{eq:defDn} \, , \\
          E_{n}(z_1,\ldots,z_{n})  &=&
 \langle \downarrow | 
  {\mathcal L}_{i}(z_1,\ldots, z_{n} ) \,  |\uparrow \rangle
\label{eq:defEn} \, .
 \end{eqnarray}
   The  operators  $D_n$,  $E_n$  and  ${Q}_n$ are $2^n \times 2^n$ matrices
  acting on ${\mathcal A}^{\otimes n}$ with scalar elements.
  We shall now prove some recursion relations satisfied by
  $D_n$, $E_n$ and $Q_n$. For $n =1$,  we have
 \begin{equation} 
  D_{1}(z_1)  =   \left( \begin{array}{cc}
                 1 & 0  \\
                 0 & z_1
                         \end{array}
                    \right)     \, , \,\, 
 E_{1}(z_1)  =   \left( \begin{array}{cc}
                 0 & 1 - z_1  \\
                 0 & 0
                         \end{array}
                    \right)     \,  \hbox { and } \, 
 Q_{1}  =   \left( \begin{array}{cc}
                 0 & 0  \\
                 1 & 0
                         \end{array}
                    \right)     \, .
\label{eq:D1E1}
 \end{equation}
    From  equations~(\ref{eq:ABA4}) and  (\ref{eq:defDn}) 
 we obtain:
\begin{equation}
         D_{n+1}(z_1,\ldots,z_{n+1})  =
  \langle \uparrow | {\mathcal L}_{i}(z_{n+1})
   \otimes {\mathcal L}_{i}(z_1,\ldots,z_n)   |\uparrow \rangle   \, . 
\label{eq:rec0D}
\end{equation}
 Using the  following identities 
  (deduced  from equations~(\ref{eq:defaL}--\ref{eq:defdL})) 
 \begin{eqnarray}
   \langle \uparrow | a (z_{n+1}) &=& \langle \uparrow | \,\,\,\, , 
\,\,\,\,\,\,\,\,\,\,\,   \langle \uparrow | b (z_{n+1}) = 0 \, ,  \\ 
  \langle \uparrow | c (z_{n+1})  &=& \langle \downarrow |  \,\,\,\,  , 
  \,\,\,\,\,\,\,\,\,\,\,
 \langle \uparrow | d (z_{n+1}) = z_{n+1}  \langle \uparrow |  \, , 
\end{eqnarray}
  we derive the following  recursion relation~:
\begin{eqnarray}
         D_{n+1}(z_1,\ldots,z_{n+1})  &=&
    \left( \begin{array}{cc} 1 & 0 \\  0 & z_{n+1}  \end{array}
                    \right)  \,  \otimes   D_n  
 +   \left( \begin{array}{cc} 0  & 0 \\  1  & 0  \end{array}
                    \right)  \,  \otimes   E_n  
 \label{eq:rec1D}
 \\  &=& \left( \begin{array}{cc}
         D_n (z_1,\ldots,z_n)  &  0 \\
         E_n (z_1,\ldots,z_n)  & z_{n+1}  D_n (z_1,\ldots,z_n)
                   \end{array}
                    \right)  \, . 
\label{eq:recD}
\end{eqnarray}
 Thus, $ D_{n+1}$ is a  $2^{n+1} \times 2^{n+1}$ matrix 
 written as  a  $2 \times 2$ matrix built  of blocks
 of size $2^{n} \times 2^{n}$.  
Similarly, we have 
\begin{eqnarray}
         E_{n+1}(z_1,\ldots,z_{n+1})   &=&
    \left( \begin{array}{cc} 0  &  1 -   z_{n+1} \\ 0  & 0 \end{array}
                    \right)  \,  \otimes   D_n 
 +   \left( \begin{array}{cc} 0  & 0 \\  0  & 1  \end{array}
                    \right)  \,  \otimes   E_n 
 \label{eq:rec1E}
 \\ &=&  \left( \begin{array}{cc}
         0   &  ( 1 - z_{n+1})    D_n (z_1,\ldots,z_n)   \\
          0   &          E_n (z_1,\ldots,z_n) 
                   \end{array}
                   \right)  \, . 
\label{eq:recE}
\end{eqnarray}

 These recursion relations, together with equation~(\ref{eq:D1E1}),
  allow to calculate   $D_n$ and  $E_n$ for all $n$.
 In particular,
 we remark that $D_n$ is a lower triangular matrix and its
 $2^n$ eigenvalues are given  by $ \prod_{k=1}^n z_k^{\epsilon_k}$,
 with $\epsilon_k = 0$ or 1. 

  Finally, using equation~(\ref{eq:defQn}), we obtain 
 \begin{equation}
 Q_{n+1}  =   \left( \begin{array}{cc}
                 0 & 0  \\
                 Q_{n} & 0
                         \end{array}
                    \right)     \, .
\label{eq:recQ}
 \end{equation}

  We now prove that the  operators  $D_n$,  $E_n$  and $Q_n$ 
 provide a   Matrix Product  Representation for 
  the components  of the eigenvector
 $|  z_1, \ldots ,  z_n    \rangle $  of the Markov matrix.  Using
 equation~(\ref{eq:Ctau}), 
   the components of  $| z_1, \ldots ,  z_n    \rangle $  on 
 a  configuration $C$ of the system can be written  as,
 \begin{equation}
  \langle C |  z_1, \ldots ,  z_n    \rangle = 
  {\rm Tr} \Big(  {Q}_n  
 \prod_{i=1}^L ( \tau_i  D_n  + (1 -\tau_i)  E_n )   \Big) \, , 
\label{eq:MPA}
\end{equation}
  where $\tau_i =1$ if $i$ is occupied in configuration $C$
   and   $\tau_i =0$ otherwise. 
  Hence, a  particle is represented by the matrix $D_n$
 and a hole is represented by the matrix $E_n$.

 Because of the  conservation of the number of particles
 and  holes,  the right hand side of equation~(\ref{eq:MPA})
 vanishes when  the configuration $C$ does not have exactly $n$ holes.
  If we call
 $x_1, x_2, \ldots, x_n$  the positions
 of the $n$ holes in  $C$,    equation~(\ref{eq:MPA}) can be rewritten as 
 \begin{equation}
  \langle x_1, \ldots ,  x_n   |  z_1, \ldots ,  z_n    \rangle = 
  {\rm Tr} \Big(  {Q}_n D_n^{x_1-1}  E_n
  D_n^{x_2-x_1 -1}  E_n  \ldots 
   D_n^{x_n-x_{n-1} -1}   E_n  D_n^{ L -x_n}  \Big) \, .
\label{eq:MPA2}
\end{equation}
This expression  gives   a   Matrix Product Representation 
 for any   eigenvector of the Markov matrix. This expression  generalizes
 the steady-state  Matrix  Product  introduced by Derrida et al. (1993).
 In the Appendix, we show that the expression~(\ref{eq:MPA2}) can be 
 recast  in the familiar  coordinate Bethe Ansatz form.

\section{Identification of the Quadratic Algebra}
\label{sec:quad}

 In the previous  section, we have derived 
 the Matrix Product Representation  from   the Algebraic Bethe Ansatz 
  by constructing  explicitly  the   operators $D_n$
  and $E_n$. We now  prove that 
  these   operators  satisfy some  simple  algebraic  relations; 
   more precisely, the  operator $E_n$ can be decomposed
 as a sum of $n$   operators $E_n^{(i)}$ 
\begin{equation}
    E_n =  \sum_{i=1}^n  E_n^{(i)} \, ,
\label{eq:decE}
\end{equation}
   that obey the quadratic relations 
 \begin{eqnarray}
       E_n^{(i)} D_n &=&  z_i  D_n  E_n^{(i)} \, , \label{eq:EiD} \\
  {\rm  for  }  \,\,\,  i \neq j~:  \,\,\,\,
    \left(1 -\frac{1}{z_i} \right)  E_n^{(i)} E_n^{(j)} 
  &=&  - \left(1 -\frac{1}{z_j} \right)     E_n^{(j)}  E_n^{(i)} 
    \label{eq:EiEj}  \, ,\\
     {\rm  and } \,\,\,   E_n^{(i)} E_n^{(i)}     &=& 0   \, .
  \label{eq:EiEi}
 \end{eqnarray} 
 In equations~(\ref{eq:EiD}--\ref{eq:EiEi}), the scalars $z_i$
 are arbitrary complex numbers and do not have to be solutions
 of the Bethe equations.  Such a   quadratic algebra
  was postulated  by Alcaraz and Lazo (2004) as an Ansatz 
   to diagonalize the  Hamiltonian of quantum  spin chains.
  We show here that, for  the ASEP,  this quadratic algebra
 can be rigorously deduced from the Algebraic Bethe Ansatz.
  Our construction is explicit and provides finite dimensional
 representations of the abstract  quadratic  algebra
  defined by  relations~(\ref{eq:EiD}--\ref{eq:EiEi}).

  We shall prove the relations~(\ref{eq:decE}--\ref{eq:EiEi})  
by induction on $n$, 
   by  diagonalizing  
  the  matrix  $D_n$ and  calculating    $E_n$ in the  new basis. We
  recall that all the matrices with index $n$ are of size 
 $2^n \times 2^n$.

 For $n =1$, the matrix   $D_1$  is diagonal, and  $D_1$ and $E_1$
 satisfy the relations~(\ref{eq:decE}--\ref{eq:EiEi}). Now, 
 we suppose, by the recursion hypothesis, 
 that we have already  diagonalized the matrix  $D_n$, {\it i.e.}, 
 we have found an invertible matrix $R_n$ 
 such that 
\begin{equation}
        R_n^{-1} D_n R_n = \Delta_n \, , 
\label{eq:defRn}
\end{equation}
where  $\Delta_n$ is a diagonal matrix with diagonal 
\begin{equation}
  {\rm diag}(\Delta_n) = 
 (1, z_1, z_2,  z_2 z_1, z_3, z_3 z_1, z_3 z_2, z_3 z_2 z_1, \ldots,
 z_n \ldots   z_1) \, . 
 \label{eq:diagDelta}
\end{equation}
 (For $n =1$, we have $\Delta_1 = D_1$  and $R_1 = 1$). 
In the  new  basis, the matrix  $E_n$ becomes 
\begin{equation}
      {\mathcal E}_n =   R_n^{-1} E_n R_n    \, . 
\label{eq:defEncal}
\end{equation}
  We  suppose, again  by the recursion hypothesis,
 that we have found a decomposition  of 
 ${\mathcal E}_n$
\begin{equation}
       {\mathcal E}_n =  \sum_{i=1}^n {\mathcal E}_n^{(i)} \, , 
\label{eq:decEncal}
\end{equation}
  such that the relations~(\ref{eq:EiD}--\ref{eq:EiEi}) are satisfied
 between $ \Delta_n $  and the  ${\mathcal E}_n^{(i)}$'s. Then, 
 by a change of basis,  the same relations are also satisfied  
 by $D_n = R_n \Delta_n R_n^{-1}$ and  $E_n^{(i)} =
 R_n  {\mathcal E}_n^{(i)} R_n^{-1}. $

 We now show that a similar  decomposition can be found  at
 the level $n+1$.   We first  construct
   the matrix $R_{n+1}$ that  transforms  $D_{n+1}$ into  the   diagonal
   form~(\ref{eq:diagDelta}). Using  equation~(\ref{eq:recD}),
  we take   $R_{n+1}$ to  be of the form
\begin{equation}
         R_{n+1} =  \left( \begin{array}{cc}
           R_n   &  0   \\
           R_n A_n  &    R_n    
                   \end{array}
                    \right)  \, , 
\label{eq:recR}
\end{equation}
where $ A_n $ is an unknown matrix to be determined. From
 equations~(\ref{eq:recD}, \ref{eq:defRn}, \ref{eq:defEncal}  
 and  \ref{eq:recR}),    we obtain
 \begin{equation}
   R_{n+1}^{-1}  D_{n+1}  R_{n+1} =  
      \left( \begin{array}{cc}
          \Delta_n   &  0   \\
   -A_n\Delta_n +  z_{n+1}\Delta_n A_n
    +  {\mathcal E}_n   &    z_{n+1}\Delta_n   
                   \end{array}
                    \right)  \, .
 \label{eq:recdiag}
  \end{equation}
 This matrix is diagonal if and  only if $ A_n $ satisfies the relation
 \begin{equation}
      A_n\Delta_n -  z_{n+1}\Delta_n A_n =  {\mathcal E}_n  \, .
 \label{eq:An}
  \end{equation}
  Knowing   that $\Delta_n$ and ${\mathcal E}_n^{(i)}$
  satisfy the relations~(\ref{eq:EiD}--\ref{eq:EiEi}), we find 
   the solution  $A_n$  of equation~(\ref{eq:An})~:
\begin{equation}
      A_n  =  \Delta_n^{-1}  \sum_{i=1}^n 
\frac{  {\mathcal E}_n^{(i)} }{ z_i -  z_{n+1}}  \, .
 \label{eq:solAn}
  \end{equation}
 We thus obtain 
 \begin{equation}
    \Delta_{n+1}  =  R_{n+1}^{-1}  D_{n+1}  R_{n+1} =  
      \left( \begin{array}{cc}
          \Delta_n   &  0   \\
       0   &    z_{n+1}\Delta_n   
                   \end{array}
                    \right)  \, ; 
 \label{eq:diagDnp1}
  \end{equation}
 $\Delta_{n+1}$  is a diagonal matrix.

The operator  $E_{n+1}$ in   the new  basis is found to be~:   
\begin{equation}
 {\mathcal E}_{n+1} =   R_{n+1}^{-1}  E_{n+1}  R_{n+1} =  
      \left( \begin{array}{cc}
   {\displaystyle   
  (1 - z_{n+1}) \sum_{i=1}^n \frac{ {\mathcal E}_n^{(i)}}{ z_i -  z_{n+1}}  }
   &   (1 - z_{n+1})\Delta_n   \\
  0   &   {\displaystyle   
  z_{n+1} \sum_{i=1}^n \frac{(z_i - 1)  {\mathcal E}_n^{(i)}}
 { z_i -  z_{n+1}} }
                   \end{array}
                    \right)  \, .
 \label{eq:rectransfE}
  \end{equation}
   This equation  is derived  by using equation~(\ref{eq:solAn})
   and the relations~(\ref{eq:EiD}--\ref{eq:EiEi}). 
  We emphasize  that  ${\mathcal E}_{n+1}$ is a
  strictly upper-triangular  matrix~:
   its  lower-left  elements and its diagonal  vanish identically.
   From equation~(\ref{eq:rectransfE}),
   we deduce the decomposition of ${\mathcal E}_{n+1}$ 
\begin{equation}
       {\mathcal E}_{n+1} =  \sum_{i=1}^{n+1} {\mathcal E}_{n+1}^{(i)} \, , 
\label{eq:recdecEncal}
\end{equation}
where  
\begin{eqnarray}
  \hbox{  for }  \,  i \le n \, , \,    {\mathcal E}_{n+1}^{(i)} &=&  
  \frac{ 1}{ z_i -  z_{n+1}}     \left( \begin{array}{cc}
   {\displaystyle   
 (1 - z_{n+1}) {\mathcal E}_n^{(i)} }
   &   0    \\
  0   &   {\displaystyle   
  z_{n+1} ( z_i - 1)  {\mathcal E}_n^{(i)} }
                   \end{array}
                    \right)  \ 
 \label{eq:recdecEi}       \\ 
 \hbox{ and } \,\,\,   {\mathcal E}_{n+1}^{(n+1)}  &=&   
      \left( \begin{array}{cc}
  0  &   (1 - z_{n+1})\Delta_n   \\
  0   &  0
                   \end{array}
                    \right)  \, .
 \label{eq:recdecEnp1}
  \end{eqnarray}

   Knowing  that $\Delta_n$ and ${\mathcal E}_n^{(i)}$
  satisfy the relations~(\ref{eq:EiD}--\ref{eq:EiEi}), and
  using  the explicit expressions~(\ref{eq:diagDnp1},
  \ref{eq:recdecEi}  and \ref{eq:recdecEnp1}), we find that the operators 
 $\Delta_{n+1}$ and ${\mathcal E}_{n+1}^{(i)}$ also 
 satisfy the algebraic rules~(\ref{eq:EiD}--\ref{eq:EiEi})
 for $ 1 \le i \le n+1$.  Reverting to the original basis 
 by using the matrix $R_{n+1}$, we conclude that 
 $D_{n+1}$ and $E_{n+1}$  satisfy the same relations.
 We have thus shown the existence of the quadratic 
 algebraic  relations~(\ref{eq:EiD}--\ref{eq:EiEi})
 at the level $n+1$.

  We finally  discuss the existence of relations between 
  the  operators  $D_n$ and $E_n$, 
   and the boundary  operator ${Q}_n$. Using
 the   equations~(\ref{eq:defQn}) and (\ref{eq:recD}), we obtain  
  the following relation between   $D_n$   and ${Q}_n$~:
\begin{equation}
        D_n {Q}_n  =  \left( \prod_{i=1}^n z_i \right)  
  {Q}_n   D_n  = \left( \prod_{i=1}^n z_i \right)  {Q}_n  
       \,  .    \label{eq:DQ} 
\end{equation}
  We emphasize that  in the  quadratic  algebra  that
  we have derived  from the algebraic Bethe Ansatz,  there is 
  no  algebraic  relation     between
  $ E_n {Q}_n$  and ${Q}_n E_n $. 
 Therefore, the  quadratic algebra that we have constructed
 is akin to but not identical to that  studied by  
 Alcaraz and Lazo (2004). However,    any  modification
 of the boundary relations   alters  the  properties
  of the algebra and  profoundly modifies  its representation theory.
  For example, it can be proved  (Golinelli and Mallick, in preparation) 
  that,  for  the ASEP,  
  the   algebra  defined  in  (Alcaraz and Lazo  2004)  is such that
   all its  finite dimensional  representations have vanishing traces and,  
    therefore,  can 
   not be used to construct a Matrix Product Ansatz.  In contrast,
  the algebra we have constructed here admits finite dimensional
   representations with non-zero trace and therefore allows  us to define
   a {\it bona-fide}   Matrix Product Ansatz. 
  Besides, from a physical point of view,  it is well known that  
boundary conditions
 play a crucial role  in the ASEP (see e.g., Sch\"utz  2001).

   The  algebra~(\ref{eq:EiD}--\ref{eq:EiEi}) and the boundary 
   equation~(\ref{eq:DQ}) encode the Bethe Ansatz. In  the  Appendix,
  we use these relations  to prove {\it ab initio} that the vector 
   $|  z_1, \ldots ,  z_n    \rangle$  whose  components
  are given in equation~(\ref{eq:MPA2}) is an eigenvector of
 the Markov matrix $M$  provided  the pseudo-moments $z_1$, $z_2$, $\ldots$,
   $z_n$ satisfy the Bethe equations~(\ref{eq:Bethe}).

\section{Generalization}
\label{sec:PASEP}

In this section,  we briefly  indicate how to generalize
  our results  to the partially asymmetric  exclusion process  in which
 the particles hop  to the  right and to the left with jump rates
 given by $p$ and $q$ respectively. 
 For the  partially asymmetric exclusion process the local
    update operator  $M_{i, i+1}$ is given by (we omit the identity
 operators for sake of  clarity)~:
\begin{equation}
         M_{i, i+1}  =  \left( \begin{array}{cccc}
                         0 &  0  &  0 &  0  \\
                         0 & - p &  q &  0  \\
                         0 &   p & -q &  0  \\
                         0 &  0  &  0 &  0  
                   \end{array}
                   \hspace{0.2in} \right)      \, .    
\label{eq:localMASEP}
\end{equation}
  Using this local operator,
 the construction of the  $ {\mathcal L}_{i}$  matrices and the 
 monodromy  operator   $T$  is identical to that  explained above.
 The recursion relations~(\ref{eq:recD}) and~(\ref{eq:recE}) 
 for the  $D_n$   and  $E_n$  operators, respectively,   
 are now replaced  by~:
\begin{equation}
         D_{n+1}(z_1,\ldots,z_{n+1})  =  \left( \begin{array}{cc}
         D_n (z_1,\ldots,z_n)  &  0 \\
 (1 - \frac{q}{p}z_{n+1}) E_n (z_1,\ldots,z_n) & z_{n+1} D_n (z_1,\ldots,z_n)
                   \end{array}
                    \right)  \, ,  
\label{eq:recpasepD}
\end{equation}
 and 
\begin{equation}
         E_{n+1}(z_1,\ldots,z_{n+1})  =  \left( \begin{array}{cc}
      \frac{q}{p}z_{n+1} E_n (z_1,\ldots,z_n)
  & (1 - z_{n+1}) D_n (z_1,\ldots,z_n)   \\
          0   &          E_n (z_1,\ldots,z_n) 
                   \end{array}
                   \right)  \, . 
\label{eq:recpasepE}
\end{equation}
 The operators $D_1$ and $E_1$ are identical to those defined
 in equation~(\ref{eq:D1E1}).
  
 Here also,   $E_n$ can be written 
 as a sum of $n$  operators $E_n^{(i)}$ as in equation~(\ref{eq:decE}).
 The operators   $E_n^{(i)}$ and $D_n$ generate a quadratic algebra.
 The relations~(\ref{eq:EiD}) and (\ref{eq:EiEi}) still hold good 
  but  the  equation~(\ref{eq:EiEj}) is replaced by  
\begin{equation}
  {\rm  for  }  \,\,\,  i \neq j~:  \,\,\,\,
 \left(1 -  \frac{p}{z_i} - q z_j \right)  E_n^{(i)} E_n^{(j)} 
  =   - \left(1 -\frac{p}{z_j}- q z_i \right)     E_n^{(j)}  E_n^{(i)} 
    \label{eq:EiEjpasep}  \, .
 \end{equation} 
 These algebraic relation are obtained, again,   by recursion on $n$.
 The  diagonal basis is found  by using the  transformation
 $R_n$ defined recursively in equation~(\ref{eq:recR})  where 
 the matrix $A_n$ is now given by 
\begin{equation}
      A_n  =  (1 - \frac{q}{p}z_{n+1})  \Delta_n^{-1}  \sum_{i=1}^n 
\frac{  {\mathcal E}_n^{(i)} }{ z_i -  z_{n+1}}  \, .
 \label{eq:solAn2}
  \end{equation}
 For $q = 0$, we recover the expressions given in sections \ref{sec:MPA}
 and  \ref{sec:quad}.

\section{Conclusion}

 In this work, we have shown that the components of the eigenvectors
 of the  asymmetric
 exclusion process can be written as traces over  matrix products.
  This  Matrix  Product Representation   has been constructed  from the
 Algebraic Bethe Ansatz in a systematic manner (\ref{eq:MPA2}).
 Our  method also allows to derive  the  
  algebraic relations (\ref{eq:decE}--\ref{eq:EiEi})
  satisfied by the  operators
 that represent  particles and holes. 
  The quadratic relations obtained
 are in fact  logical consequences of the  Algebraic  Bethe Ansatz procedure  
 and thus,  ultimately, stem  from the Yang-Baxter equation. 
 The approach described in this work shows that there is  a close relation
 between the  Matrix method  and the  Bethe Ansatz, at least in the case
 of the ASEP.  We believe that the derivation  of the
  Matrix Ansatz  for the ASEP  presented here  
 can be generalized   to   integrable  quantum chains. 
  Besides, if the equivalence   between  Matrix  Ansatz  and Bethe Ansatz 
 is true in general, this would
 provide a technique for constructing the  quadratic algebras
 from first principles rather that having to postulate them a priori. 
  We also emphasize that, in contrast with the 
 work of Alcaraz and Lazo (2004),   our  construction provides 
 explicit finite dimensional 
 representations of the algebras involved that can be used
 for actual calculations on finite size systems. 

  \subsection*{Acknowledgments}
    We thank Bernard Derrida, Vincent Hakim and Vincent Pasquier
 for help and encouragement  at early stages of this work. 
 We also thank S. Mallick for a careful reading of the manuscript. 

\section*{Appendix: Derivation of the Bethe Equations}

  We  explain here how to derive  the Bethe equations~(\ref{eq:Bethe})
 from the quadratic algebra~(\ref{eq:EiD}--\ref{eq:EiEi})
 and the boundary relation~(\ref{eq:DQ}). We first show that 
  the expression~(\ref{eq:MPA2}) is equivalent to 
  the standard coordinate Bethe Ansatz form:
\begin{eqnarray}
 & & \langle x_1, \ldots ,  x_n   |  z_1, \ldots ,  z_n    \rangle = 
 \nonumber \\  & & {\rm Tr} \Big(  {Q}_n D_n^{x_1-1}  E_n
  D_n^{x_2-x_1 -1}  E_n  \ldots 
   D_n^{x_n-x_{n-1} -1}   E_n  D_n^{ L -x_n}  \Big) =  \nonumber \\
   & &\sum_{\sigma \in \Sigma_n} 
   {\rm Tr} \Big(  {Q}_n D_n^{x_1-1}  E_n^{(\sigma(1))}
  D_n^{x_2-x_1 -1}  E_n^{(\sigma(2))}  \ldots 
   D_n^{x_n-x_{n-1} -1}   E_n^{(\sigma(n))}  D_n^{ L -x_n}  \Big) \, , 
 \,\,\,\, \,\,\,\, 
\label{eq:CoordBA}
\end{eqnarray}
   where  $\sigma$   belongs to  $\Sigma_n$ the permutation group
 of $n$ objects.  This formula  is obtained  by inserting
 the decomposition~(\ref{eq:decE}) for   the
 operators  $E_n$  and by noticing from 
 relation~(\ref{eq:EiEi}) that each $E_n^{(i)}$ must 
 appear only  once.  We  use equation~(\ref{eq:EiD}) to push
 all the  operators $E_n^{(\sigma(i))}$ to  the right. We thus obtain
\begin{eqnarray}
 & & \langle x_1,\ldots ,  x_n   |  z_1,\ldots ,  z_n  \rangle = \nonumber \\
    & &\sum_{\sigma \in \Sigma_n} z_{\sigma(n)}^{L-x_n}
 z_{\sigma(n-1)}^{L-1-x_{n-1}}  \ldots z_{\sigma(1)}^{L-n +1-x_1}
  {\rm Tr} \Big(  {Q}_n D_n^{L-n}  E_n^{(\sigma(1))}
  E_n^{(\sigma(2))}  \ldots 
   E_n^{(\sigma(n))}   \Big)  \,.   \,\,\,\, \,\,\,\,\,\,\,\, 
\label{eq:CoordBA2}
\end{eqnarray}
 We use  equation~(\ref{eq:EiEj}) to rearrange the  product
  $E_n^{\sigma(1)} \ldots E_n^{\sigma(n)}$ in the canonical order
 $E_n^{(1)} \ldots E_n^{(n)}$ and  obtain
\begin{equation}
  \langle x_1,\ldots ,  x_n   |  z_1,\ldots ,  z_n  \rangle =
  K  \sum_{\sigma \in \Sigma_n} (-1)^{\sigma} 
  \prod_{i=1}^n  \, \left(z_{\sigma(i)} - 1 \right)^{i} 
  z_{\sigma(1)}^{-x_1}\ldots  z_{\sigma(n)}^{-x_n} 
\end{equation}
with
\begin{equation}
  K  = \left( \prod_{i=1}^n z_i\right)^{L-n}
  \prod_{i=1}^n( 1 - \frac{1}{ z_i})^{-1}
  {\rm Tr} \Big(  {Q}_n D_n^{L-n}  E_n^{(1)}
  E_n^{(2)}  \ldots 
   E_n^{(n)}   \Big) \, , \,\,\,  
\end{equation}
   and  where $(-1)^{\sigma}$ represents the signature of the 
 permutation $\sigma$.  We  thus find   that
 the eigenvector can be written  (Golinelli and Mallick 2005) as a determinant
 of a matrix  $V_{i,j}$~:
\begin{equation}
  \langle x_1,\ldots ,  x_n   |  z_1,\ldots ,  z_n  \rangle = 
 K \det( V_{i,j} )  \,\,\, \hbox{ with }\,\,\, 
 V_{i,j} = (z_j -1)^i z_j^{-x_i}  \, .
\end{equation}

 We now  show 
 that for  the vector~(\ref{eq:MPA2}) to be  an  
   eigenvector of  the Markov matrix $M$, the 
  pseudo-moments $z_1, \ldots z_n$  must satisfy the 
 Bethe equations~(\ref{eq:Bethe}).  Using  the fact that
  an   eigenvector of  $M$  
 is also an  eigenvector of the transfer matrix
 $t(\lambda)$ for any value of $\lambda$  
  and therefore of  the translation  operator
  ${\mathcal T}= t(0),$ we obtain 
 \begin{eqnarray}
 \langle x_1,\ldots ,  x_n   |{\mathcal T}  | z_1, z_2 \ldots  z_n \rangle
  &=&   \langle x_1-1,\ldots ,  x_n-1| z_1, z_2 \ldots  z_n \rangle 
   \nonumber   \\ &=&  
   \zeta   \langle x_1,\ldots ,  x_n  | z_1, z_2 \ldots  z_n  \rangle  \, ,
 \label{eq:TranOp}  
  \end{eqnarray}
    with  $\zeta^L = 1\,.$  We now  substitute equation~(\ref{eq:MPA2})
  in  this identity. We have to distinguish two cases:  $x_1 > 1$
 and $x_1 =  1$.  For $x_1 > 1$,  we  have
 \begin{eqnarray}
   {\rm Tr} \Big(  {Q}_n D_n^{x_1-2}  E_n
  D_n^{x_2-x_1 -1}  E_n  \ldots 
   D_n^{x_n-x_{n-1} -1}   E_n  D_n^{ L -x_n} D \Big)  &=&
    \nonumber \\   \zeta \,\,    {\rm Tr} \Big(  {Q}_n D_n^{x_1-1}  E_n
  D_n^{x_2-x_1 -1}  E_n  \ldots 
   D_n^{x_n-x_{n-1} -1}   E_n  D_n^{ L -x_n} \Big)  &=&
 \nonumber \\    
 \left( \prod_{i=1}^n z_i \right)   {\rm Tr} \Big(  {Q}_n D_n^{x_1-1}  E_n
  D_n^{x_2-x_1 -1}  E_n  \ldots 
   D_n^{x_n-x_{n-1} -1}   E_n  D_n^{ L -x_n} \Big). 
\end{eqnarray}
 To derive the last equality, we have used  equation~(\ref{eq:DQ}).
  We thus have  
    \begin{equation}
  z_1 \ldots  z_n  = \zeta  \, .
  \label{eq:Moment}
 \end{equation}
  For $x_1 =  1$, equation~(\ref{eq:TranOp}) becomes
 \begin{eqnarray}
   {\rm Tr} \Big(  {Q}_n   
  D_n^{x_2-2}  E_n  \ldots 
   D_n^{x_n-x_{n-1} -1}   E_n  D_n^{ L -x_n}  E_n \Big)  &=&
    \nonumber \\   \zeta \,\,    {\rm Tr} \Big(  {Q}_n  E_n
  D_n^{x_2-2}  E_n  \ldots 
   D_n^{x_n-x_{n-1} -1}   E_n  D_n^{ L -x_n} \Big) \, .  & &
  \label{eq:OpTr2}
    \end{eqnarray} 
    Using the decomposition~(\ref{eq:CoordBA}),
 we obtain the following sufficient condition for
  equation~(\ref{eq:OpTr2}) to be satisfied  by any  $\sigma$~:
 \begin{eqnarray}
   {\rm Tr} \Big(  {Q}_n   
  D_n^{x_2-2}  E_n^{(\sigma(2))}   \ldots 
   D_n^{x_n-x_{n-1} -1}   E_n  D_n^{ L -x_n}  E_n^{(\sigma(1))}  \Big)  &=&
    \nonumber \\   \zeta \,\,    {\rm Tr} \Big(  {Q}_n  E_n^{(\sigma(1))}
  D_n^{x_2-2}  E_n^{(\sigma(2))}  \ldots 
   D_n^{x_n-x_{n-1} -1}   E_n^{(\sigma(n))}  D_n^{ L -x_n} \Big) \, .& & 
    \end{eqnarray} 
   Using   equation~(\ref{eq:EiEj}), we commute   the operator 
  $E_n^{(\sigma(1))}$ with all the other operators
  $E_n^{(\sigma(j))}$, for $ j \neq 1$, 
   and bring it   back to the rightmost position. 
 This leads to the consistency condition 
  \begin{equation}
 1 = \zeta z_{\sigma(1)} ^{L} (-1)^{n-1} \prod_{i=1}^n 
  \frac{z_i -1}{ z_i(z_{\sigma(1)} - 1)}  \, .
 \end{equation}
   From  relation~(\ref{eq:Moment}), we conclude   that   
 this equation is identical  to  the Bethe equation~(\ref{eq:Bethe}).

\section*{References}

\begin{itemize}

\item 
 Alcaraz~F.~C., Droz~M., Henkel~M., Rittenberg~V.,  1994,
{\em Reaction-diffusion processes, critical dynamics,  and quantum
 chains},
 Ann. Phys.  {\bf 230}, 250.

\item 
Alcaraz F.~C.  and Lazo M.~J. , 2004,
{\em The Bethe ansatz as a matrix product ansatz}, 
  J. Phys. A: Math. Gen. {\bf 37}, L1;
 {\it  Exact solutions of exactly integrable quantum  chains
 by a matrix product ansatz},   J. Phys. A: Math. Gen. {\bf 37}, 4149.

\item Derrida B., 1998, {\em An exactly soluble non-equilibrium
    system: the asymmetric simple exclusion process}, Phys. Rep.  {\bf
    301}, 65.

\item Derrida B., Evans M.~R., Hakim V., Pasquier V., 1993, {\em Exact
    solution of a 1D asymmetric exclusion model using a matrix
    formulation}, J. Phys. A: Math. Gen. {\bf 26}, 1493.

\item Derrida B.  and Lebowitz J.~L., 1998, {\em Exact large deviation
    function in the asymmetric exclusion process}, Phys. Rev. Lett.
  {\bf 80}, 209.

\item Derrida B., Lebowitz J.~L., Speer E.~R., 2003, {\em Exact large
    deviation functional of a stationary open driven diffusive system:
    the asymmetric exclusion process}, J. Stat. Phys. {\bf 110}, 775.

\item Dhar D., 1987, {\em An exactly solved model for interfacial
    growth}, Phase Transitions {\bf 9}, 51.

\item  Golinelli O. and    Mallick K., 2004,
 {\em  Hidden symmetries in the asymmetric
  exclusion process}, J. Stat. Mech. P12001.

\item Golinelli O. and Mallick K., 2005, 
 {\em  Spectral gap of the totally asymmetric exclusion
 process at arbitrary filling}, 
  J. Phys. A: Math. Gen.  {\bf 38}, 1419.

\item Gwa L.-H., Spohn H., 1992, {\em Bethe solution for the
    dynamical-scaling exponent of the noisy Burgers equation}, Phys.
  Rev. A {\bf 46}, 844.

\item Kim D., 1995, {\em Bethe ansatz solution for crossover scaling
    functions of the asymmetric XXZ chain and the
    Kardar-Parisi-Zhang-type growth model}, Phys. Rev. E {\bf 52},
  3512.

\item   Mallick K., 1996, {\em Syst\`emes hors d'\'equilibre:
 quelques r\'esultats exacts}, PhD Thesis, University of Paris 6.  
  
\item Nepomechie R.I., 1999,  {\it A Spin Chain Primer},
 Int. J. Mod. Phys. B  {\bf 13}, 2973  (hep-th/9810032).

\item Sch\"utz G.~M., 2001, {\it Phase Transitions and Critical Phenomena,}
  Vol. 19, C.~Domb and J.~L.~Lebowitz Eds, (Academic, London).

\item 
 Speer E.~R., 1993,
{\em The two species totally  asymmetric exclusion  process},
 in Micro, Meso and Macroscopic approaches in Physics, M.~Fannes
 et al.  Eds. (NATO Workshop, Leuven, 1993).

\item 
  Stinchcombe R.~B. and Sch\"utz  G.~M., 1995,
 {\em  Application of operator algebras  to stochastic dynamics
  and the Heisenberg chain}, 
 Phys. Rev. Lett.  {\bf 75}, 140.

\end{itemize}

\end{document}